\documentclass[10pt,conference,letterpaper]{IEEEtran}

\pdfoutput=1

\ifCLASSINFOpdf
\else
\fi
\usepackage{float}
\usepackage{graphicx}
\usepackage{subfigure}
\usepackage{multicol}
\usepackage{algorithm}
\usepackage{algpseudocode}
\usepackage{amsmath}
\usepackage{amssymb}
\usepackage{mathrsfs}
\usepackage{amsthm}
\usepackage{bm}
\usepackage{cite}
\usepackage{enumitem}
\usepackage{diagbox}
\usepackage{color}

\newcommand{\cK}{\mathcal{K}}
\newcommand{\cP}{\mathcal{P}}
\newcommand{\cN}{\mathcal{N}}

\newcommand{\cU}{\mathcal{U}}

\newcommand{\bw}{\boldsymbol w}
\newcommand{\bp}{\boldsymbol p}

\newcommand{\bx}{\boldsymbol x}

\newcommand{\bP}{\boldsymbol P}
\newcommand{\bzero}{\boldsymbol 0}

\newtheorem{myth}{Theorem}

\newtheorem{myle}[myth]{Lemma}

\IEEEoverridecommandlockouts

\allowdisplaybreaks

\begin{document}
\title{Weighted Sum-Rate Maximization in Multi-Carrier NOMA with Cellular Power Constraint}

\setlength{\skip\footins}{0.25cm}

\author{
\thanks{A part of the work was carried out at LINCS (www.lincs.fr).}%
\IEEEauthorblockN{Lou Sala\"un\IEEEauthorrefmark{1}\IEEEauthorrefmark{2}, Marceau~Coupechoux\IEEEauthorrefmark{2} and Chung~Shue~Chen\IEEEauthorrefmark{1}}
\IEEEauthorblockA{\IEEEauthorrefmark{1}Bell Labs, Nokia Paris-Saclay, 91620 Nozay, France}
\IEEEauthorblockA{\IEEEauthorrefmark{2}LTCI, Telecom ParisTech, University of Paris-Saclay, France}
\IEEEauthorblockA{Email: lou.salaun@nokia-bell-labs.com, marceau.coupechoux@telecom-paristech.fr, chung\_shue.chen@nokia-bell-labs.com}
}

\maketitle

\begin{abstract}
Non-orthogonal multiple access (NOMA) has received significant attention for future wireless networks. NOMA outperforms orthogonal schemes, such as OFDMA, in terms of spectral efficiency and massive connectivity. 
The joint subcarrier and power allocation problem in NOMA is NP-hard to solve in general, due to complex impacts of signal superposition on each user's achievable data rates, as well as combinatorial constraints on the number of multiplexed users per sub-carrier to mitigate error propagation.
In this family of problems, weighted sum-rate (WSR) is an important objective function as it can achieve different tradeoffs between sum-rate performance and user fairness.
We propose a novel approach to solve the WSR maximization problem in multi-carrier NOMA with cellular power constraint. The problem is divided into two polynomial time solvable sub-problems. First, the multi-carrier power control (given a fixed subcarrier allocation) is non-convex. By taking advantage of its separability property, we design an optimal and low complexity algorithm (\textsc{MCPC}) based on projected gradient descent. Secondly, the single-carrier user selection is a non-convex mixed-integer problem that we solve using dynamic programming (\textsc{SCUS}). 
This work also aims to give an understanding on how each sub-problem's particular structure can facilitate the algorithm design. In that respect, the above \textsc{MCPC} and \textsc{SCUS} are basic building blocks that can be applied in a wide range of resource allocation problems. Furthermore, we propose an efficient heuristic to solve the general WSR maximization problem by combining \textsc{MCPC} and \textsc{SCUS}.
Numerical results show that it achieves near-optimal sum-rate with user fairness, as well as significant performance improvement over OMA.
\end{abstract}

\vspace{-0.1cm}

\section{Introduction}\label{sec:intro}

In multi-carrier multiple access systems, the total frequency bandwidth is divided into subcarriers and assigned to users to optimize the spectrum utilization. Orthogonal multiple access (OMA), such as orthogonal frequency-division multiple access (OFDMA) adopted in 3GPP-Long Term Evolution 4G standards~\cite{dahlman20134g}, only serves one user per subcarrier in order to avoid intra-cell interference and have low-complexity signal decoding at the receiver side. OMA is known to be suboptimal in terms of spectral efficiency~\cite{cover2012elements}.

The principle of multi-carrier non-orthogonal multiple access (MC-NOMA) is to multiplex several users on the same subcarrier by performing signals superposition at the transmitter side. Successive interference cancellation (SIC) is applied at the receiver side to mitigate interference between superposed signals. MC-NOMA is a promising multiple access technology for 5G mobile networks as it achieves better spectral efficiency and higher data rates than OMA schemes~\cite{saito2013non,dai2015non,oviedo2016new}. 

Careful optimization of the transmit powers is required to control the intra-carrier interference of superposed signals and maximize the achievable data rates. 
Besides, due to error propagation and decoding complexity concerns in practice~\cite{tse2005fundamentals}, subcarrier allocation for each transmission also needs to be optimized. 
As a consequence, joint subcarrier and power allocation problems in NOMA have received much attention. In this class of problems, weighted sum-rate (WSR) maximization is especially important as it can achieve different tradeoffs between sum-rate performance and user fairness~\cite{weeraddana2012weighted}. 
One may also consider to use the weights to perform fair resource allocation in a stable scheduling~\cite{eryilmaz2007fair,tassiulas1992stability}.

Two types of power constraints are considered in the literature. On the one hand, cellular power constraint is mostly used in downlink transmissions to represent the total transmit power budget available at the base station (BS). On the other hand, individual power constraint sets a power limit independently for each user. The latter is clearly more appropriate for uplink scenarios~\cite{chen2015suboptimal,al2014uplink}, nevertheless it can also be applied to the downlink~\cite{lei2016power,fu2018subcarrier}. 

It is known that WSR maximization in OFDMA systems is polynomial time solvable if we consider cellular power constraint~\cite{liu2014complexity1} but strongly NP-hard with individual power constraints~\cite{liu2014complexity2}. 
It has also been proven that MC-NOMA with individual power constraints is strongly NP-hard~\cite{lei2016power,salaun2018optimal}. 
Several algorithms are developed to perform subcarrier and/or power allocation in this setting. Fractional transmit power control (FTPC) is a simple heuristic that allocates a fraction of the total power budget to each user based on their channel condition~\cite{ding2016impact}. 
In~\cite{chen2015suboptimal} and~\cite{al2014uplink}, heuristic user pairing strategies and iterative resource allocation algorithms are studied for uplink transmissions.
A time efficient iterative waterfilling heuristic is introduced in~\cite{fu2017double} to solve the problem with equal weights.  
Reference~\cite{lei2016power} derives an upper bound on the optimal WSR and proposes a near-optimal scheme using dynamic programming and Lagrangian duality techniques. This scheme serves as benchmark due to its high computational complexity, which may not be suitable for practical systems with low latency requirements. 

To the best of our knowledge, it is not known if MC-NOMA with cellular power constraint is polynomial time solvable or NP-hard. 
Only a few papers have developed optimization schemes for this problem, which are either heuristics with no theoretical performance guarantee or algorithms with impractical computational complexity.
For example, a greedy user selection and heuristic power allocation scheme based on difference-of-convex programming is proposed in~\cite{parida2014power}. The authors of~\cite{sun2016optimal} employ monotonic optimization to develop an optimal resource allocation policy, which serves as benchmark due to its exponential complexity. The algorithm of~\cite{lei2016power} can also be applied to cellular power constraint scenarios, but it has very high complexity as well. 

Motivated by this observation, we investigate the WSR maximization problem in a downlink multi-carrier NOMA system with cellular power constraint. Our contributions are as follows:
\begin{enumerate}[leftmargin=*]
\item We propose a novel framework in which the original problem is divided into two polynomial time solvable sub-problems:
\begin{itemize}
\item The multi-carrier power control (given a fixed subcarrier allocation) sub-problem is non-convex. By taking advantage of its separability property, we design an optimal and low complexity algorithm (\textsc{MCPC}) based on projected gradient descent. 
\item The single-carrier user selection sub-problem is a non-convex mixed-integer problem that we solve using dynamic programming (\textsc{SCUS}). 
\end{itemize}
This work also aims to give an understanding on how each sub-problem's particular structure can facilitate the algorithm design. In that respect, the above \textsc{MCPC} and \textsc{SCUS} are basic building blocks that can be applied in a wide range of resource allocation problems. Our analysis of the separability property in \textsc{MCPC} and the combinatorial constraints in \textsc{SCUS} provide mathematical tools which can be used to study other similar resource allocation optimization problems. 
\item Papers~\cite{parida2014power,sun2016optimal} restrict the number of multiplexed users per subcarrier, denoted by $M$, to be equal to 2. However, our result can be applied for an arbitrary number of superposed signals on each subcarrier, i.e., $M \geq 1$. The value of $M$ can be set depending on error propagation and decoding complexity concerns in practice.
\item We propose an efficient heuristic to solve the general WSR maximization problem by combining \textsc{MCPC} and \textsc{SCUS}.
Numerical results show that it achieves near-optimal sum-rate and user fairness, as well as significant improvement in performance over OMA.
\end{enumerate}

The paper is organized as follows. In Section~\ref{sec:model}, we present the system model and notations. Section~\ref{sec:pbl} formulates the WSR problem. We introduce our algorithms in Section~\ref{sec:algo} and analyze their optimality and complexity. We show in Section~\ref{sec:num} some numerical results, highlighting our solution's sum-rate and fairness performance, as well as their computational complexity. Finally, we conclude in Section~\ref{sec:ccl}.

\section{System Model and Notations}\label{sec:model}
We define in this section the system model and notations used throughout the paper. We consider a downlink multi-carrier NOMA system composed of one base station (BS) serving $K$ users. We denote the index set of users by $\cK \triangleq \{1,\ldots,K\}$, and the set of subcarriers by ${\cN \triangleq \{1,\ldots,N\}}$. The total system bandwidth $W$ is divided into $N$ subcarriers of bandwidth $W_n = W/N$, for each $n\in\cN$. We assume orthogonal frequency division, so that adjacent subcarriers do not interfere each other. Moreover, each subcarrier $n\in\cN$ experiences frequency-flat block fading on its bandwidth $W_n$.

Let $p_k^n$ denotes the transmit power from the BS to user $k\in\cK$ on subcarrier $n\in\cN$. User $k$ is said to be active on subcarrier $n$ if $p_k^n > 0$, and inactive otherwise. In addition, let $g_k^n$ be the channel gain between the BS and user $k$ on subcarrier $n$, and $\eta_k^n$ be the received noise power. For simplicity of notations, we define the normalized noise power as $\tilde{\eta}_k^n \triangleq \eta_k^n/g_k^n$. We denote by ${\bp \triangleq\left(p_k^n\right)_{k\in\cK,n\in\cN}}$ the vector of all transmit powers, and ${\bp^n \triangleq\left(p_k^n\right)_{k\in\cK}}$ the vector of transmit powers on subcarrier $n$.

In power domain NOMA, several users are multiplexed on the same subcarrier using superposition coding. A common approach adopted in the literature is to limit the number of superposed signals on each subcarrier to be no more than $M$. The value of $M$ is meant to characterize practical limitations of SIC due to decoding complexity and error propagation~\cite{tse2005fundamentals}.
We represent the set of active users on subcarrier $n$ by ${\cU_n \triangleq \{k\in\cK \colon p_k^n >0\}}$. The aforementioned constraint can then be formulated as $\forall n\in\cN,\; |\cU_n| \leq M$, where $|\mathord{\cdot}|$ denotes the cardinality of a finite set. Each subcarrier is modeled as a multi-user Gaussian broadcast channel~\cite{tse2005fundamentals} and SIC is applied at the receiver side to mitigate intra-band interference.

The SIC decoding order on subcarrier $n$ is usually defined as a permutation function over the active users on $n$, i.e., $\pi_n \colon \{1,\ldots,|\cU_n|\} \to \cU_n$. However, for ease of reading, we choose to represent it by a permutation over all users $\cK$, i.e., $\pi_n \colon \{1,\ldots,K\} \to \cK$. These two definitions are equivalent in our model since the Shannon capacity~\eqref{eq:R_2} does not depend on the inactive users $k\in\cK\setminus\cU_n$, for which $p_k^n = 0$. For $i\in\{1,\ldots,K\}$, $\pi_n(i)$ returns the $i$-th decoded user's index. Conversely, user $k$'s decoding order is given by $\pi_n^{-1}(k)$. 

In this work, we consider the optimal decoding order studied in~\cite[Section 6.2]{tse2005fundamentals}. It consists of decoding users' signals from the highest to the lowest normalized noise power:
\begin{equation}\label{SIC_DL}
\tilde{\eta}_{\pi_n(1)}^n \geq \tilde{\eta}_{\pi_n(2)}^n \geq \cdots \geq \tilde{\eta}_{\pi_n(K)}^n.
\end{equation}
User $\pi_n(i)$ first decodes the signals of users $\pi_n(1)$ to $\pi_n(i-1)$ and subtracts them from the superposed signal before decoding its own signal. Interference from users $\pi_n(j)$ for $j > i$ is treated as noise. 
The maximum achievable data rate of user $k$ on subcarrier $n$ is given by Shannon capacity:
\begin{align} 
R_k^n(\bp^n) &\triangleq W_n\log_2\left(1+\frac{g_k^n p_k^n}{\sum_{j=\pi^{-1}_n(k)+1}^{K}{g_k^n p_{\pi_n(j)}^n}+\eta^n_k}\right) \nonumber\\
&\stackrel{(a)}{=} W_n\log_2\left(1+\frac{p_k^n}{\sum_{j=\pi^{-1}_n(k)+1}^{K}{p_{\pi_n(j)}^n}+\tilde{\eta}^n_k}\right), \label{eq:R_2}
\end{align}
where equality $(a)$ is obtained after normalizing by $g_k^n$. We assume perfect SIC, therefore interference from users $\pi_n(j)$ for $j < \pi^{-1}_n(k)$ is completely removed in~\eqref{eq:R_2}. 
 
\section{Problem Formulation}\label{sec:pbl}

Let ${\bw = \{w_1,\ldots,w_K\}}$ be a sequence of $K$ positive weights. The main focus of this work is to solve the following joint subcarrier and power allocation optimization problem:

\begin{equation}\tag{$\cP$}\label{P}
\begin{aligned}
& \underset{\bp}{\text{maximize}}
& & \sum_{k\in\cK}{w_k\sum_{n\in\cN}{R_k^n\left(\bp^n\right)}}, \\
& \text{subject to}
& & C1:~\sum_{k\in\cK}\sum_{n\in\cN}{p_k^n} \leq P_{max}, \\
&&& C2:~\sum_{k\in\cK}{p_k^n} \leq P_{max}^n,~n\in\cN, \\
&&& C3:~p_k^n \geq 0,~k\in\cK,~n\in\cN, \\
&&& C4:~|\cU_n| \leq M,~n\in\cN. \\
\end{aligned}
\end{equation}
The objective of~\ref{P} is to maximize the system's WSR. As discussed in Section~\ref{sec:intro}, this objective function has received much attention since its weights $\bw$ can be chosen to achieve different tradeoffs between sum-rate performance and fairness~\cite{weeraddana2012weighted}. The weights can also be tuned to perform fair resource allocation in a stable scheduling~\cite{eryilmaz2007fair,tassiulas1992stability}. Note that $C1$ represents the cellular power constraint, i.e., a total power budget $P_{max}$ at the BS. In $C2$, we set a power limit of $P_{max}^n$ for each subcarrier $n$. This is a common assumption in multi-carrier systems, e.g.,~\cite{liu2014complexity1,liu2014complexity2}. Constraint $C3$ ensures that the allocated powers remain non-negative. Due to decoding complexity and error propagation in SIC~\cite{tse2005fundamentals}, we restrict the number of multiplexed users per subcarrier to $M$ in $C4$. 

Let us consider the following change of variables:
\begin{equation}\label{eq:change}
\forall n\in\cN,\; x_i^n \triangleq \begin{cases}
    \sum_{j=i}^{K}{p_{\pi_n(j)}^n}, & \text{if $i\in\{1,\ldots,K\}$},\\
    0, & \text{if $i=K+1$}.
  \end{cases} 
\end{equation}
We define ${\bx \triangleq\left(x_i^n\right)_{i\in\{1,\ldots,K\},n\in\cN}}$ and ${\bx^n \triangleq\left(x_i^n\right)_{i\in\{1,\ldots,K\}}}$.
\begin{myle}[Equivalent problem~\ref{P2}]\label{le:equivalence}
$ $\newline
Problem~\ref{P} is equivalent to problem~\ref{P2} formulated below:
\begin{align}
& \underset{\bx}{\text{maximize}}
& & f(\bx) = \sum_{n\in\cN}{\sum_{i=1}^{K}{f_i^n(x_i^n)}}, \tag{$\cP'$}\label{P2}\\
& \text{subject to}
& & C1':~\sum_{n\in\cN}{x_1^n} \leq P_{max}, \nonumber\\
&&& C2':~x_1^n \leq P_{max}^n,~n\in\cN, \nonumber\\
&&& C3':~x_i^n \geq x_{i+1}^n,~i\in\{1,\ldots,K\},~n\in\cN, \nonumber\\
&&& C3'':~x_{K+1}^n=0,~n\in\cN, \nonumber\\
&&& C4':~|\cU'_n| \leq M,~n\in\cN, \nonumber
\end{align}
where for any $i\in\{1,\ldots,K\}$ and $n\in\cN$, we have:
\begin{equation*}
f_i^n(x_i^n) \triangleq \begin{cases}
W_n\log_2\left(\left(x_1^n+\tilde{\eta}^n_{\pi_n(1)}\right)^{w_{\pi_n(1)}}\right), & \text{if $i=1$}, \\
W_n\log_2\left(\frac{\left(x_i^n+\tilde{\eta}^n_{\pi_n(i)}\right)^{w_{\pi_n(i)}}}{\left(x_i^n+\tilde{\eta}^n_{\pi_n(i-1)}\right)^{w_{\pi_n(i-1)}}}\right), & \text{if $i > 1$},
\end{cases}
\end{equation*}
and where ${\cU'_n \triangleq \{i\in\{1,\ldots,K\} \colon x_i^n > x_{i+1}^n\}}$.
\end{myle}
\begin{IEEEproof}
See Appendix~\ref{ap:transf}.
\end{IEEEproof}

The advantage of this formulation~\ref{P2} is that it exhibits a separable objective function in both dimensions $i\in\{1,\ldots,K\}$ and $n\in\cN$. In other words, it can be written as a sum of functions $f_i^n$, each only depending on one variable $x_i^n$. In the following section, we take advantage of this separability property to design efficient algorithms to solve~\ref{P2}. The solution of~\ref{P} can then be obtained by solving~\ref{P2}.

\section{Algorithm Design}\label{sec:algo}
Decomposing the joint subcarrier and power allocation problem~\ref{P2} into smaller sub-problems allows us to develop efficient algorithms by taking advantage of each sub-problem's particular structure. The technical details are given below.


\subsection{Single-Carrier User Selection}\label{sec:algo_B}

In this subsection, we restrict the optimization search space to a single subcarrier. Given $n\in\cN$ and a power budget $\bar{P}^n$, the single-carrier user selection sub-problem~\ref{Psub2} consists in finding a power allocation on $n$ satisfying the multiplexing and SIC constraint $C4'$.
\begin{align}
& \underset{\bx^n}{\text{maximize}}
& & {\sum_{i=1}^{K}{f_i^n(x_i^n)}}, \tag{$\cP'_{SCUS}(n)$}\label{Psub2}\\
& \text{subject to}
& & C2', C3', C3'', C4'. \nonumber
\end{align}

Let us introduce auxiliary functions that will help us in our algorithm design. For $n\in\cN$, $i\in\{1,\ldots,K\}$ and $j \leq i$, assume that the consecutive variables $x_j^n,\ldots,x_i^n$ are all equal to a certain value $x\in\left[0,\bar{P}^n\right]$. We define $f_{j,i}^n$ as:
\begin{align}
f_{j,i}^n(x) &\triangleq \sum_{l=j}^{i}{f_l^n(x)} \nonumber\\
&= \begin{cases}
W_n\log_2\left(\left(x+\tilde{\eta}^n_{\pi_n(i)}\right)^{w_{\pi_n(i)}}\right), & \text{if $j=1$}, \\
W_n\log_2\left(\frac{\left(x+\tilde{\eta}^n_{\pi_n(i)}\right)^{w_{\pi_n(i)}}}{\left(x+\tilde{\eta}^n_{\pi_n(j-1)}\right)^{w_{\pi_n(j-1)}}}\right), & \text{if $j > 1$}.\nonumber
\end{cases}
\end{align}
This simplification of notation is relevant for the analysis of \textsc{SCUS} (Algorithm~\ref{alg3}) and also the coming \textsc{MCPC} (Algorithm~\ref{algoMCPC}). Indeed, if users $j,\ldots,i-1$ are not active (i.e., $j,\ldots,i-1\notin\cU'_n$), then $x_j^n = \cdots = x_i^n$, therefore $\sum_{l=j}^{i}{f_l^n}$ can be replaced by $f_{j,i}^n$ and $x_{j+1}^n , \ldots , x_i^n$ are redundant with $x_j^n$. If constraint $C4'$ is satisfied, up to $M$ users are active on each subcarrier. Thus, evaluating and optimizing the objective function of~\ref{Psub2} only requires $O(M)$ operations. 

\begin{figure}[!ht]
\vspace{-0.5cm}
\centering
\includegraphics[width=0.85\linewidth]{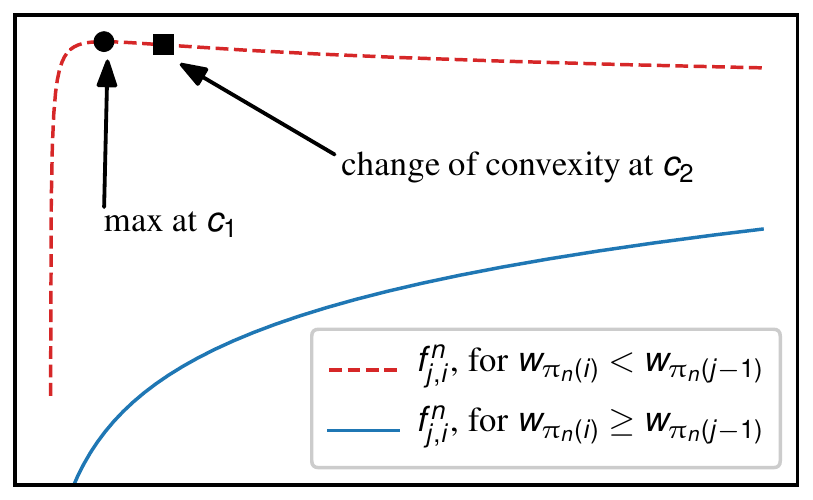}
\vspace{-0.2cm}
\caption{The two general forms of functions $f_{j,i}^n$}
\label{fig_f}
\vspace{-0.3cm}
\end{figure}

We study the properties of $f_{j,i}^n$ in Lemma~\ref{le:maxfij}. Note that $f_i^n = f_{i,i}^n$, therefore Lemma~\ref{le:maxfij} also holds for functions $f_i^n$. Fig.~\ref{fig_f} shows an example of $f_{j,i}^n$.

\begin{myle}[Properties of $f_{j,i}^n$]\label{le:maxfij}
$ $\newline
Let $n\in\cN$, $i\in\{1,\ldots,K\}$, and $j \leq i$, we have:
\begin{itemize}
\item If $j=1$ or $w_{\pi_n(i)} \geq w_{\pi_n(j-1)}$, then $f_{j,i}^n$ is strictly increasing and concave on $\left[0,\infty\right)$.
\item Otherwise when $j>1$ and $w_{\pi_n(i)} < w_{\pi_n(j-1)}$, $f_{j,i}^n$ is strongly unimodal. It increases on $\left(-\tilde{\eta}_{\pi_n(j-1)},c_1\right]$ and decreases on $\left[c_1,\infty\right)$, where
\begin{equation*}
c_1 = \frac{w_{\pi_n(j-1)}\tilde{\eta}_{\pi_n(i)}-w_{\pi_n(i)}\tilde{\eta}_{\pi_n(j-1)}}{w_{\pi_n(i)}-w_{\pi_n(j-1)}}.
\end{equation*}
Besides, $f_{j,i}^n$ is concave on $\left(-\tilde{\eta}_{\pi_n(j-1)},c_2\right]$ and convex on $\left[c_2,\infty\right)$, where
\begin{equation*}
c_2 = \frac{\sqrt{w_{\pi_n(j-1)}}\tilde{\eta}_{\pi_n(i)}-\sqrt{w_{\pi_n(i)}}\tilde{\eta}_{\pi_n(j-1)}}{\sqrt{w_{\pi_n(i)}}-\sqrt{w_{\pi_n(j-1)}}} \geq c_1.
\end{equation*}
\end{itemize}
\end{myle}
\begin{IEEEproof}
The idea consists in studying the first and second derivatives of $f_{j,i}^n$. Details are omitted due to space limitation. 
\end{IEEEproof}

\algnewcommand{\LineComment}[1]{\State \(\triangleright\) #1}
\newcommand{\algruleinput}[1][.2pt]{\par\vskip.2\baselineskip\hrule height #1\par\vskip.5\baselineskip}
\algnewcommand\algorithmicinput{\textbf{Input:}}
\algnewcommand\Input{\item[\algorithmicinput]}
\algnewcommand\algorithmicfunc{\textbf{function}}
\algnewcommand\MyFunc{\item[\algorithmicfunc]}
\algnewcommand\algorithmicendfunc{\textbf{end function }}
\algnewcommand\EndFunc{\item[\algorithmicendfunc]}


We present in Algorithm~\ref{algo:MaxF} the pseudocode \textsc{MaxF} which computes the maximum of $f_{j,i}^n$ on $\left[0,\bar{P}^n\right]$ following the result of Lemma~\ref{le:maxfij}. \textsc{MaxF} only requires a constant number of basic operations, therefore its complexity is $O(1)$. For ease of reading, we summarize some system parameters of a given instance of~\ref{P}, for all $n\in\cN$, as follows:
\begin{equation*}\label{param}
\mathcal{I}^n = \left(\bw,\cK,W_n,(g_k^n)_{k\in\cK},(\eta_k^n)_{k\in\cK}\right).
\end{equation*}

\vspace{-0.2cm}
\begin{algorithm}[h]
\caption{Compute maximum of $f_{j,i}^n$ on $\left[0,\bar{P}^n\right]$}\label{algo:MaxF}
\begin{algorithmic}[1]
\MyFunc \textsc{MaxF}($j,i, \mathcal{I}^n, \bar{P}^n$)
\State $a \gets \pi_n(i)$
\State $b \gets \pi_n(j-1)$
\If{$j=1$ or $w_a \geq w_b$}
\State \Return $\bar{P}^n$
\Else
\State \Return $\max\left\{0,\min\left\{\frac{w_b\tilde{\eta}_a-w_a\tilde{\eta}_b}{w_a-w_b},\bar{P}^n\right\}\right\}$
\EndIf
\EndFunc
\end{algorithmic}
\end{algorithm}

We propose to solve~\ref{Psub2} using Algorithm~\ref{alg3} (\textsc{SCUS}) based on dynamic programming (DP). This approach consists in computing recursively the elements of three arrays $V$, $X$, $T$. Let $m\in\{0,\ldots,M\},\;j\in\{1,\ldots,K\}$ and $i\geq j$, we define $V[m,j,i]$ as the optimal value of~\ref{Psub2} with no more than $m$ users multiplexed on subcarrier $n$ in constraint $C4'$, and with the additional constraints $C5'$ and $C6'$ as follows:
\begin{align}
V[m,j,i] & & \triangleq &&& \underset{\bx^n}{\max} & & {\sum_{i=1}^{K}{f_i^n(x_i^n)}}, \tag{$\cP'_{SCUS}[m,j,i]$}\label{defV}\\ 
&&&&& \text{subject to} & & C2', C3', C3'',\nonumber\\
&&&&&&& C4':~|\cU'_n| \leq m, \nonumber\\
&&&&&&& C5':~x_j^n = \cdots = x_i^n, \nonumber\\
&&&&&&& C6':~x_l^n = 0, l > i. \nonumber
\end{align}

\newlength{\textfloatsepsave} 
\setlength{\textfloatsepsave}{\textfloatsep} 
\setlength{\textfloatsep}{0.10cm}
\begin{algorithm}[!t]\small
\caption{Single-carrier user selection algorithm (\textsc{SCUS})}\label{alg3}
\begin{algorithmic}[1]
\MyFunc \textsc{SCUS}($\mathcal{I}^n, \bar{P}^n, M$)
\LineComment Initialize arrays $V$, $X$, $U$ for $m=0$ and $j=0$
\For{$j = 0$ to $K$ and $i = j$ to $K$}
\State $V\left[0,j,i\right] \gets 0$
\State $X\left[0,j,i\right] \gets 0$
\State $U\left[0,j,i\right] \gets (0,0,0)$
\EndFor
\For{$m = 1$ to $M$ and $i = 0$ to $K$}
\State $V\left[m,0,i\right] \gets 0$
\State $X\left[m,0,i\right] \gets 0$
\State $U\left[m,0,i\right] \gets (0,0,0)$
\EndFor
\LineComment Compute $V$, $X$, $U$ for $m\leq M$ and $j\leq i \leq K$
\For{$j = 1$ to $K$ and $i = j$ to $K$ and $m = 1$ to $M$}
\State $x^* \gets \textsc{MaxF}(j,i, \mathcal{I}^n,\bar{P}^n)$
\State $v_0 \gets V[m,j-1,j-1]$
\State $v_1 \gets V[m-1,j-1,j-1] + f_{j,i}^n\left(x^*\right) - f_{j,i}^n\left(0\right)$
\State $v_2 \gets V[m,j-1,i]$
\If{$0 < x^* < X\left[m-1,j-1,j-1\right]$ and $v_1 \geq v_0$ and $v_1 \geq v_2$}
\State $V\left[m,j,i\right] \gets v_1$
\State $X\left[m,j,i\right] \gets x^*$
\State $U\left[m,j,i\right] \gets (m-1,j-1,j-1)$
\ElsIf{$v_2 \geq v_0$}
\State $V\left[m,j,i\right] \gets v_2$
\State $X\left[m,j,i\right] \gets X\left[m,j-1,i\right]$
\State $U\left[m,j,i\right] \gets (m,j-1,i)$
\Else
\State $V\left[m,j,i\right] \gets v_0$
\State $X\left[m,j,i\right] \gets 0$
\State $U\left[m,j,i\right] \gets (m,j-1,j-1)$
\EndIf
\EndFor
\LineComment Retrieve the optimal solution $\bx^n$
\State $x_1^n,\ldots,x_K^n \gets 0$
\State $(m,j,i) \gets (M,K,K)$
\Repeat 
\State $x_j^n,\ldots,x_i^n \gets X\left[m,j,i\right]$
\State $(m,j,i) \gets U\left[m,j,i\right]$
\Until{$(m,j,i) = (0,0,0)$}
\State\Return $\bx^n$
\EndFunc
\end{algorithmic}
\end{algorithm}

It is interesting to note that $V[M,K,K]$ is the optimal value of~\ref{Psub2}, since constraints $C5'$ and $C6'$ become trivially true for $j = i= K$. Let ${x_1^n}^*,\ldots,{x_K^n}^*$ be the optimal solution achieving $V[m,j,i]$. We define $X[m,j,i] \triangleq {x_i^n}^*$, which is also equal to ${x_j^n}^*,\ldots,{x_{i-1}^n}^*$ due to constraint $C5'$. 

The idea of SCUS is to recursively compute the elements of $V$ for $m\in\{0,\ldots,M\},\;j\in\{1,\ldots,K\}$ and $i\in\{j\ldots,K\}$ through the following recurrence relation:
\begin{equation}\label{SCUSrec}
V[m,j,i] = \max\{v_0,v_1,v_2\},
\end{equation}
\vspace{-0.6cm}
\begin{flalign*}
v_0 &= V[m,j-1,j-1], \nonumber\\
v_1 &= 
\begin{cases}
&V[m-1,j-1,j-1] + f_{j,i}^n\left(x^*\right) - f_{j,i}^n\left(0\right),\\
& \hphantom{0,} \quad\quad\quad\quad \text{ if } 0 < x^* < X\left[m-1,j-1,j-1\right],\\
&0, \quad\quad\quad\quad \text{ otherwise},
\end{cases}\nonumber\\
v_2 &= V[m,j-1,i],\nonumber
\end{flalign*}
where $x^* = \textsc{MaxF}(j,i, \mathcal{I}^n, \bar{P}^n)$, and $v_0$, $v_1$, $v_2$ correspond to allocations where user $i$ and $j-1$ are each either active or inactive. A detailed analysis is given in Appendix~\ref{ap:thalgoSCUS}.

During SCUS's iterations, the array $U$ keeps track of which previous element of $V$ (i.e., $V[m,j-1,j-1]$ or $V[m-1,j-1,j-1]$ or $V[m,j-1,i]$) have been used to compute the current value function $V[m,j,i]$. More precisely, $U[m,j,i]$ contains one of the following indices $(m,j-1,j-1)$, $(m-1,j-1,j-1)$ or $(m,j-1,i)$. This allows us to retrieve the entire optimal vector $\bx^n$ at the end of the algorithm (at lines 32-38) by backtracking from index $(M,K,K)$ to $(0,0,0)$. 

When $m = 0$, no user can be active on this subcarrier due to constraint $C4'$. Therefore, $V$, $X$, $U$ can be initialized to zero at lines~1 to~6. For simplicity, we also extend $V$, $X$ and $U$ on the index $j = 0$ and $i \geq 0$ and initialize them to zero in lines~7 to~11.

\begin{myth}[Optimality and complexity of algorithm \textsc{SCUS}]\label{th:algoSCUS}
Given a subcarrier $n\in\cN$, a power budget $\bar{P}^n$ and $M\geq 1$, algorithm \textsc{SCUS} computes the optimal single-carrier power control and user selection of~\ref{Psub2}. Its worst case computational complexity is $O(MK^2)$.
\end{myth}
\begin{IEEEproof}
See Appendix~\ref{ap:thalgoSCUS}.
\end{IEEEproof}

\subsection{Multi-Carrier Power Control}\label{sec:algo_A}

\begin{figure}[!ht]
\vspace{-0.5cm}
\centering
\includegraphics[width=0.95\linewidth]{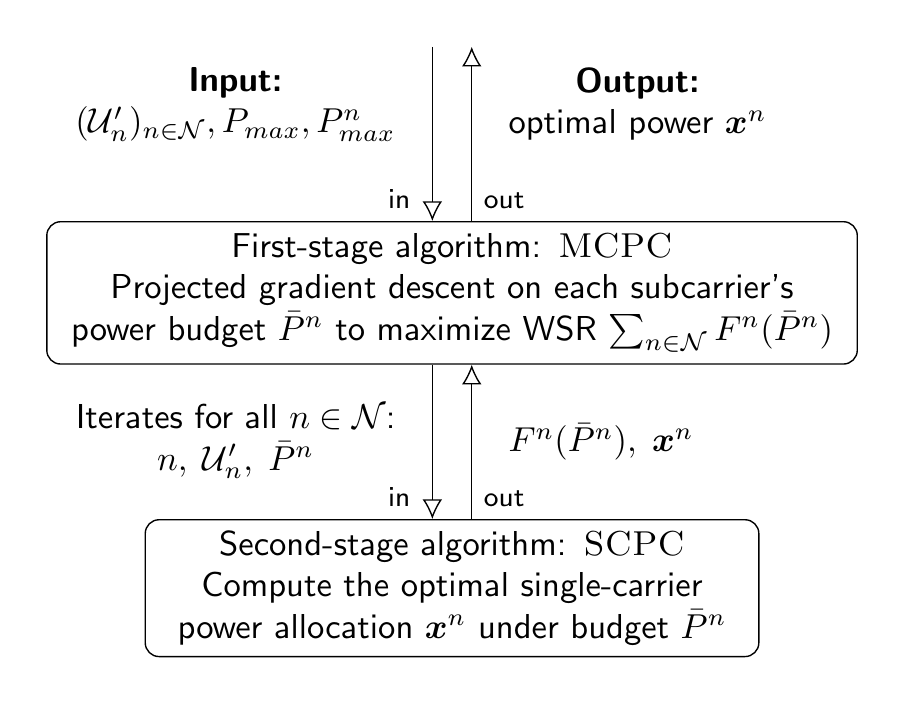}
\vspace{-0.5cm}
\caption{Overview of \textsc{MCPC}}
\label{fig_diagraMCPC}
\vspace{-0.1cm}
\end{figure}

The following multi-carrier power control sub-problem~\ref{Psub1} consists in maximizing the WSR taking as input a fixed subcarrier allocation $\cU'_n$, for all $n\in\cN$, so that constraint $C4'$ can be ignored. Since inactive users $k\notin\cU_n$ on each subcarrier $n\in\cN$ have no contribution on the data rates, i.e., $p_k^n = 0$ and $R_k^n = 0$, we remove them in the definition of~\ref{P} for the study of this sub-problem. Without loss of generality, we then transform~\ref{P} to~\ref{P2} and index the remaining active users on each subcarrier $n$ by $i_n\in\{1_n,\ldots,|\cU'_n|_n\}$. We formulate~\ref{Psub1} as a two-stage optimization problem.  The first-stage is:
\begin{align}
& \underset{\bar{\bP}}{\text{maximize}}
& & \sum_{n\in\cN}F^n(\bar{P}^n) \tag{$\cP'_{MCPC}$}\label{Psub1}\\
& \text{subject to}
& & \bar{P}^n \in \mathcal{F}, \nonumber
\end{align}
where $\bar{P}^n$, for $n\in\cN$, are intermediate variables representing each subcarrier's power budget. $\bar{\bP} \triangleq \left(\bar{P}^1,\ldots,\bar{P}^N\right)$ denotes the power budget vector. The feasible set 
\begin{equation}\label{eq:feasibleSet}
\mathcal{F} \triangleq \{\bar{\bP} \colon \sum_{n\in\cN}\bar{P}^n \leq P_{max} \text{ and }  0 \leq \bar{P}^n \leq P_{max}^n,\; n\in\cN\}
\end{equation} 
is chosen to satisfy $C1'$ and $C2'$. $F^n(\bar{P}^n)$ is the optimal value of the second-stage:
\begin{align}
F^n(\bar{P}^n) & & = &&& \underset{x_{i_n}^n}{\max} & & \sum_{i=1}^{|\cU'_n|}{f_{i_n}^n(x_{i_n}^n)} \tag{$\cP'_{SCPC}$}\label{PsubSCPC}\\
&&&&& \text{subject to} & & x_{1_n}^n \leq \bar{P}^n,\nonumber\\
&&&&&&& x_{1_n}^n \geq x_{(i+1)_n}^n,~i\in\{1,\ldots,|\cU'_n|\}, \nonumber\\
&&&&&&& x_{(|\cU'_n|+1)_n}^n=0. \nonumber
\end{align}
The first-stage~\ref{Psub1} consists in optimizing the allocated power budget $\bar{P}^n$ among subcarriers $n\in\cN$. While the second-stage~\ref{PsubSCPC} performs power allocation on each subcarrier $n$, with a given allocation $\cU'_n$ and power budget $\bar{P}^n$. Fig.~\ref{fig_diagraMCPC} gives an overview of the nested optimization that we design to solve~\ref{Psub1}. 

We solve the second-stage~\ref{PsubSCPC} using Algorithm~\ref{algoSCPC} (\textsc{SCPC}).
The idea is to iterate over variables $x_{i_n}^n$ for $i = 1$ to $|\cU'_n|$, and compute their optimal value $x^* = \textsc{MaxF}({i_n},{i_n}, \mathcal{I}^n,\bar{P}^n)$ at line~3. If the current allocation satisfies constraint $C3'$, then $x_{i_n}^n$ gets value $x^*$. Otherwise, the algorithm backtracks at line~6 and finds the highest index $j \in\{2,\ldots,i-1\}$ such that $x_{(j-1)_n}^n \geq \textsc{MaxF}({j_n},{i_n}, \mathcal{I}^n,\bar{P}^n)$. Then, variables $x_{j_n}^n,\ldots,x_{i_n}^n$ are set equal to $\textsc{MaxF}({j_n},{i_n}, \mathcal{I}^n,\bar{P}^n)$ at line~10. The optimality and complexity of this algorithm are presented in Theorem~\ref{th:algoSCPC}.

\begin{algorithm}[t]
\caption{Single-carrier power control algorithm (\textsc{SCPC})}\label{algoSCPC}
\begin{algorithmic}[1]
\MyFunc \textsc{SCPC}($\mathcal{I}^n, \cU'_n, \bar{P}^n$)
\For{$i=1$ to $|\cU'_n|$}
\LineComment Compute the optimal of $f_{i_n}^n$
\State $x^* \gets$ \textsc{MaxF}(${i_n},{i_n}, \mathcal{I}^n,\bar{P}^n$)
\LineComment Modify $x^*$ if this allocation violates constraint $C3'$
\State $j \gets i-1$
\While{$x_{j_n}^n < x^*$ and $j\geq 1$}
\State $x^* \gets$ \textsc{MaxF}(${j_n},{i_n}, \mathcal{I}^n,\bar{P}^n$)
\State $j \gets j-1$
\EndWhile
\State $x_{(j+1)_n}^n,\ldots,x_{i_n}^n \gets x^*$
\EndFor
\State \Return $x_{1_n}^n,\ldots,x_{|\cU'_n|_n}^n$
\EndFunc
\end{algorithmic}
\end{algorithm}

\begin{myth}[Optimality and complexity of algorithm \textsc{SCPC}]\label{th:algoSCPC}
Given subcarrier $n\in\cN$ and a power budget $\bar{P}^n$, algorithm \textsc{SCPC} computes the optimal single-carrier power control. Its worst case computational complexity is $O(|\cU'_n|^2)$.
\end{myth}
\begin{IEEEproof}
See Appendix~\ref{ap:thalgoSCPC}.
\end{IEEEproof}


\begin{myle}[Convexity of~\ref{Psub1}]\label{lemma:Fconcave}
$\mathcal{F}$ is a convex set and $\sum_{n\in\cN}F^n$ is concave with respect to $\bar{\bP}\in\mathcal{F}$.
\end{myle}
\begin{IEEEproof}
See Appendix~\ref{ap:Fconcave}.
\end{IEEEproof}

According to Lemma~\ref{lemma:Fconcave}, $\sum_{n\in\cN}F^n$ is concave on the convex feasible set $\mathcal{F}$. Hence, the optimal multi-carrier power allocation~\ref{Psub1} can be computed efficiently using projected gradient descent~\cite{boyd2004convex}. This \textsc{MCPC} scheme is described in Algorithm~\ref{algoMCPC}, in which the second-stage optimal value $F^n(\bar{P}^n)$, defined in~\ref{PsubSCPC}, is determined using \textsc{SCPC} as described in the auxiliary function in lines~12 to~13. The search direction at line~4 can be found either using numerical methods or an exact gradient formula (in the Proof of Lemma~\ref{lemma:Fconcave}). Note that the step size $\alpha$ at line~5 can be tuned by backtracking line search or exact line search~\cite{boyd2004convex}. We adopt the latter to perform simulations. The projection of $\bar{\bP} + \alpha\Delta$ on the simplex $\mathcal{F}$ at line~6 can be computed efficiently~\cite{calamai1987projected}, the details of its implementation are omitted here.

\begin{algorithm}[t]
\caption{Multi-carrier power control algorithm (\textsc{MCPC})}\label{algoMCPC}
\begin{algorithmic}[1]
\MyFunc \textsc{MCPC}$\left(\left(\mathcal{I}^n\right)_{n\in\cN},(\cU'_n)_{n\in\cN},P_{max},P^n_{max},\cN,\epsilon\right)$
\State Let $\bar{\bP} \gets \bzero$ be the starting point
\Repeat
\State Save the previous vector $\bar{\bP}' \gets \bar{\bP}$
\State Determine a search direction $\Delta \gets \nabla \sum_{n\in\cN}{F^n(\bar{P}^n)}$
\State Choose a step size $\alpha$
\State Update $\bar{\bP} \gets$ projection of $\bar{\bP} + \alpha\Delta$ on $\mathcal{F}$
\Until{$||\bar{\bP}'-\bar{\bP}||_2^2\leq \epsilon$}
\For{$n\in\cN$}
\State $x_{1_n}^n,\ldots,x_{|\cU'_n|_n}^n \gets \textsc{SCPC}\left(\mathcal{I}^n, \cU'_n, \bar{P}^n\right)$
\EndFor
\State \Return $(x_{1_n}^n,\ldots,x_{|\cU'_n|_n}^n)_{n\in\cN}$
\EndFunc

\MyFunc \algruleinput $F^n\left(\bar{P}^n\right) \triangleq F^n\left(\mathcal{I}^n, \cU'_n, \bar{P}^n\right)$
\State $x_{1_n}^n,\ldots,x_{|\cU'_n|_n}^n \gets \textsc{SCPC}\left(\mathcal{I}^n, \cU'_n, \bar{P}^n\right)$
\State \Return $\sum_{i=1}^{|\cU'_n|}{f_{i_n}^n(x_{i_n}^n)}$
\EndFunc
\end{algorithmic}
\end{algorithm}

\begin{myth}[Optimality and complexity of algorithm \textsc{MCPC}]\label{th:algoMCPC}
Given a subcarrier allocation such that $|\cU'_n| \leq M$, for all $n\in\cN$, algorithm \textsc{MCPC} converges to the optimal solution of~\ref{Psub1} in $O(\log(1/\epsilon))$ iterations, where $\epsilon$ is the error tolerance. Each iteration requires $O(NM^2)$ basic operations. 
\end{myth}
\begin{IEEEproof}
It follows from Lemma~\ref{lemma:Fconcave} and classical convex programming results~\cite{boyd2004convex} that projected gradient descent converges to the optimal multi-carrier power control (\textsc{MCPC}) in $O(\log(1/\epsilon))$ iterations. Each iteration requires to compute \textsc{SCPC}($\mathcal{I}^n, \cU'_n, \bar{P}^n$), for $n\in\cN$. Thus, we derive from Theorem~\ref{th:algoSCPC} and $|\cU'_n| \leq M$ that each iteration's worst case complexity is $O(NM^2)$.
\end{IEEEproof}

\subsection{Multi-Carrier Joint Subcarrier and Power Allocation}\label{sec:algo_C}
\begin{figure}[t!]
\vspace{-0.65cm}
\centering
\includegraphics[width=0.945\linewidth]{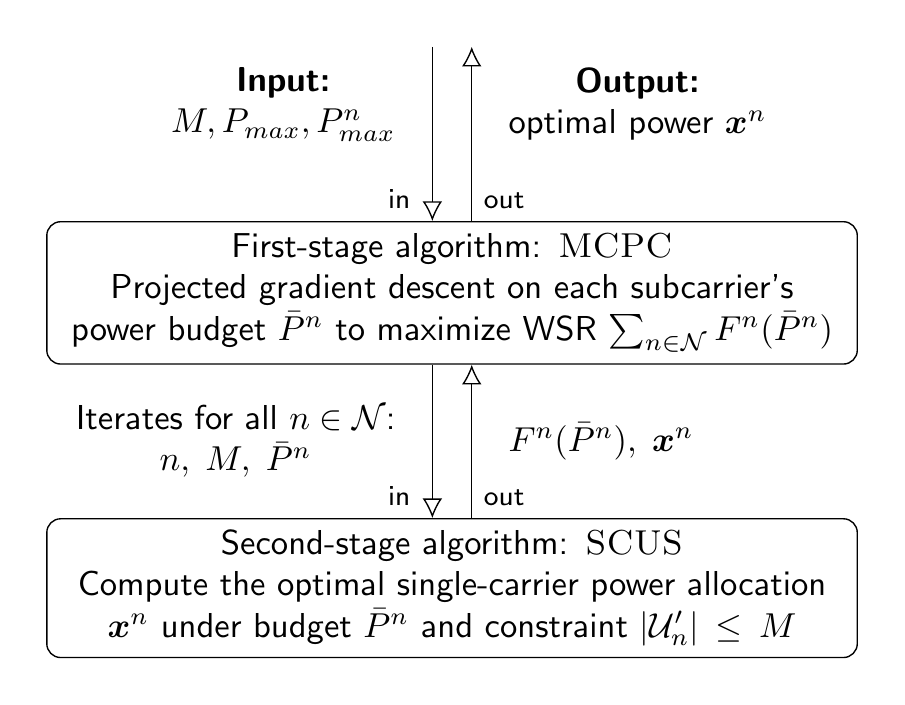}
\vspace{-0.6cm}
\caption{Overview of \textsc{JSPA}}
\label{fig_diagraJSPA}
\vspace{-0.1cm}
\end{figure}

We design an efficient heuristic joint subcarrier and power allocation scheme, namely \textsc{JSPA}, which combines \textsc{MCPC} and \textsc{SCUS}. We present its principle in Fig.~\ref{fig_diagraJSPA}.
\textsc{JSPA} is similar to the two-stage \textsc{MCPC} depicted in Fig.~\ref{fig_diagraMCPC}. With the difference that subcarrier allocation is no longer provided as input, and has to be optimized in the second-stage. To this end, we replace the second-stage \textsc{SCPC} by the optimal single-carrier user selection algorithm \textsc{SCUS} studied in Section~\ref{sec:algo_B}.  Although \textsc{SCUS} is optimal, the returned $F^n(\bar{P}^n)$ is no longer concave. As a consequence, \textsc{JSPA} is not guaranteed to converge to a global maximum. Each iteration of \textsc{JSPA} requires to compute \textsc{SCUS}($\mathcal{I}^n, \bar{P}^n, M$), for $n\in\cN$. Thus, we derive from Theorem~\ref{th:algoSCUS} that each iteration's worst case complexity is $O(NMK^2)$. Nevertheless, we show by numerical results in the next section that it achieves near-optimal weighted sum-rate performance with low complexity. 


\section{Numerical Results}\label{sec:num}
We evaluate the WSR, user fairness performance and computational complexity of \textsc{JSPA} through numerical simulations. We compare our proposed scheme with the near-optimal high complexity benchmark scheme \textsc{LDDP} introduced in~\cite{lei2016power}, as well as FTPC with greedy subcarrier allocation, which is also considered in~\cite{lei2016power} and~\cite{ding2016impact,fu2017double,parida2014power}.
We consider a cell of diameter $500$ meters, with one BS located at its center and $K$ users distributed uniformly at random in the cell. The number of users $K$ varies from $5$ to $30$, and the number of subcarriers is $N = 10$. Each point in the following figures are average value obtained over $1000$ random instances. The simulation parameters and channel model are summarized in Table~\ref{parameter}. 

\begin{figure}[!t]
\vspace{-0.5cm}
\centering
\includegraphics[width=0.94\linewidth]{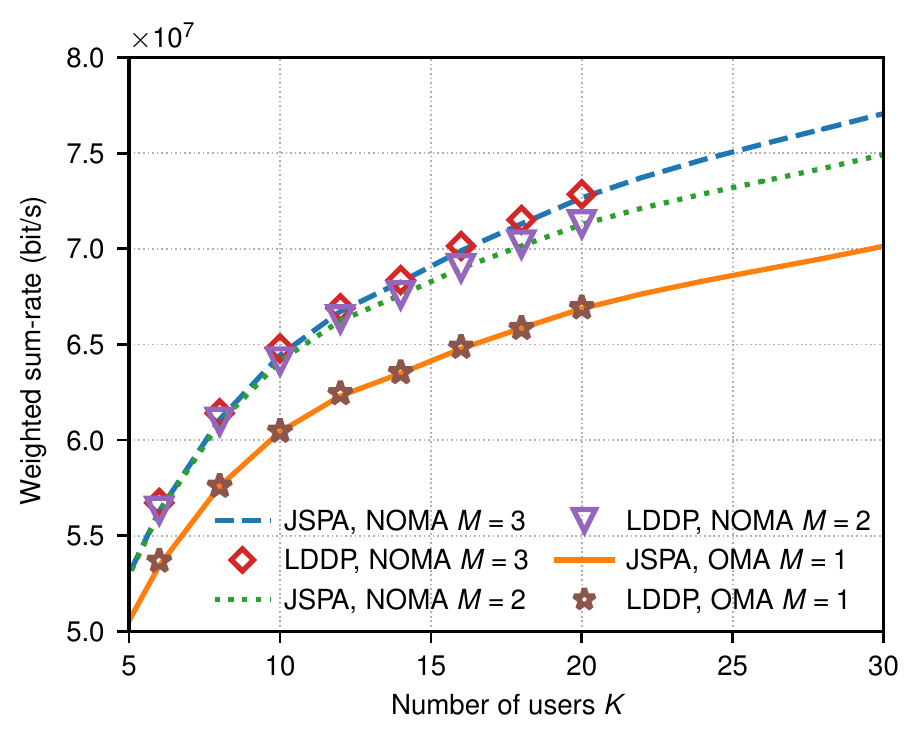}
\vspace{-0.4cm}
\caption{WSR versus $K$ for \textsc{JSPA} and \textsc{LDDP}}
\label{fig_wsr1}
\end{figure}

\begin{figure}[!t]
\vspace{-0.5cm}
\centering
\includegraphics[width=0.94\linewidth]{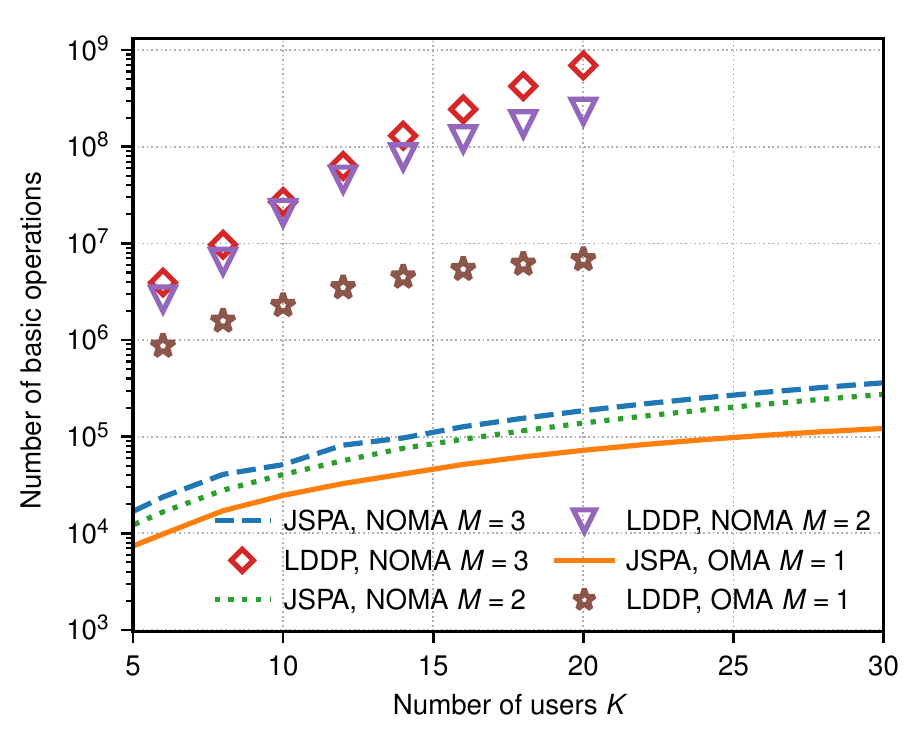}
\vspace{-0.4cm}
\caption{Number of basic operations versus $K$ for \textsc{JSPA} and \textsc{LDDP}}
\label{fig_cplx}
\vspace{-0cm}
\end{figure}

Fig.~\ref{fig_wsr1} and~\ref{fig_cplx} illustrate the WSR and complexity of \textsc{JSPA} and \textsc{LDDP} for different systems, i.e., OMA with $M=1$, NOMA with $M=2$ and $M=3$. In these two figures, weights $\bw$ are generated uniformly at random in $\left[0,1\right]$. Algorithm \textsc{LDDP} requires to discretize the total power budget $P_{max}$ into $J$ power levels. Here, we choose $J=200$. Due to the high computational complexity of \textsc{LDDP}, we only simulate it for $K=5$ to $20$.

In Fig.~\ref{fig_wsr1}, the performance loss of \textsc{JSPA} compared to \textsc{LDDP} is less than $0.8\%$ for any number of users $K$. This indicates that \textsc{JSPA} also achieves near-optimal WSR, close to that of \textsc{LDDP}. In addition, we see that the performance gain of NOMA systems compared to OMA is $7\%$ for $M=2$ and $10\%$ for $M=3$.

\setlength{\skip\footins}{0.25cm}

In Fig.~\ref{fig_cplx}, we count the number of basic operations (additions, multiplications, comparisons) performed by each scheme, which reflects their computational complexity. As expected from Theorem~\ref{th:algoSCPC}, the complexity of \textsc{JSPA} increases linearly with $M$ (there is a constant difference between the curves of \textsc{JSPA} with $M=1$, $2$ and $3$ in the semi-log plot). 
In addition, each \textsc{LDDP}'s iteration has complexity $O(NMKJ^2)$~\cite{lei2016power}. Reference~\cite{fu2018subcarrier} shows that the number of power levels should be $J = \Theta(\min\{K,NM\})$ to achieve near-optimal WSR. Thus, \textsc{JSPA} has lower asymptotic complexity than \textsc{LDDP}. For example with $K \leq NM$, LDDP requires $J = \Theta(K)$ power levels, and \textsc{JSPA} is asymptotically faster by a factor $K$. In our simulations, \textsc{JSPA} runs within seconds on a common computer\footnote{with the following specifications: Python 3, Windows 7, 64bits, AMD A10-5750M APU with Radeon HD Graphics, and 8 GB of RAM.} for $K \leq 30$, whereas \textsc{LDDP} requires 1600 times more operations for $K=20$ and $M=2$. 

\begin{table}[!t]
\vspace{-0.2cm}
\begin{center}
\caption{Simulation parameters}
\label{parameter}
\scalebox{0.94}{
\begin{tabular}{|c|c|}
  \hline
  \textbf{Parameters}  & \textbf{Value} \\ \hline
  Cell radius  & 250 m \\ \hline
  Minimum distance from user to BS  & 35 m \\ \hline
	Carrier frequency & $2$ GHz \\ \hline
  Path loss model & $128.1 + 37.6 \log_{10} d$ dB, $d$ is in km \\ \hline
  Shadowing  & Log-normal, $8$ dB standard deviation \\ \hline
  Fading & Rayleigh fading with variance 1 \\ \hline
  Noise power spectral density & -174 dBm/Hz \\ \hline
  System bandwidth $W$ & $5$ MHz \\ \hline
  Number of subcarriers $N$ & $10$  \\ \hline
  Number of users $K$ & $5$ to $30$  \\ \hline
	Total power budget $P_{max}$ & 1 W \\ \hline
  Error tolerance $\epsilon$ & $10^{-4}$\\ \hline
  Parameter $M$ &1 (OMA), 2 and 3 (NOMA)\\
  \hline
\end{tabular}}
\end{center}
\vspace{-0.1cm}
\end{table}

\begin{figure}[!t]
\vspace{-0.5cm}
\centering
\includegraphics[width=0.94\linewidth]{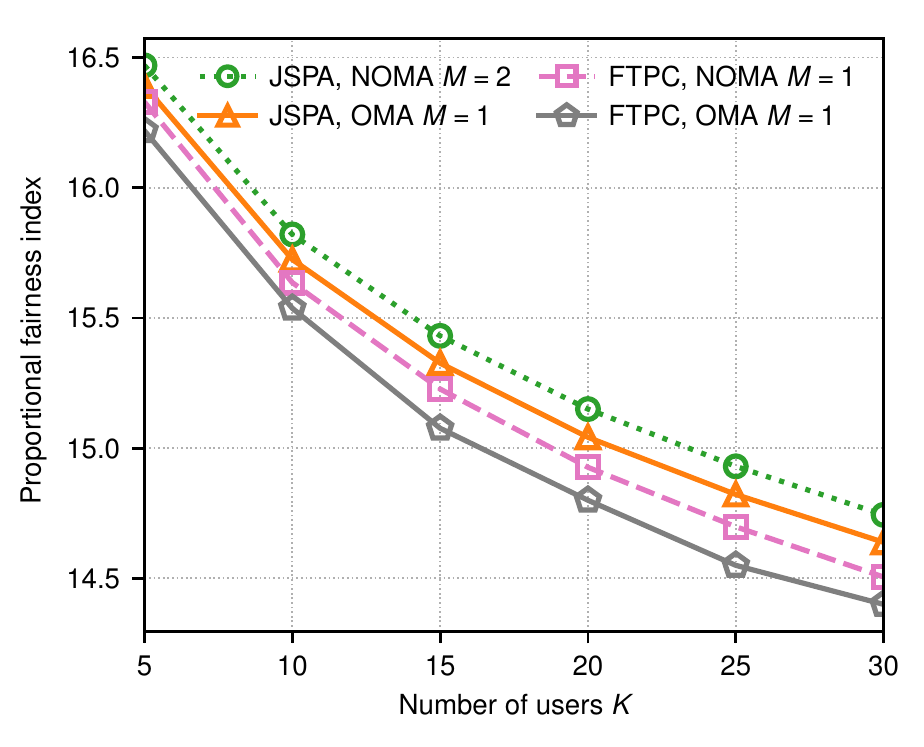}
\vspace{-0.5cm}
\caption{Proportional fairness index of \textsc{JSPA} and \textsc{FTPC}}
\label{fig_PFI}
\vspace{-0cm}
\end{figure}

\begin{figure}[!t]
\vspace{-0.5cm}
\centering
\includegraphics[width=0.94\linewidth]{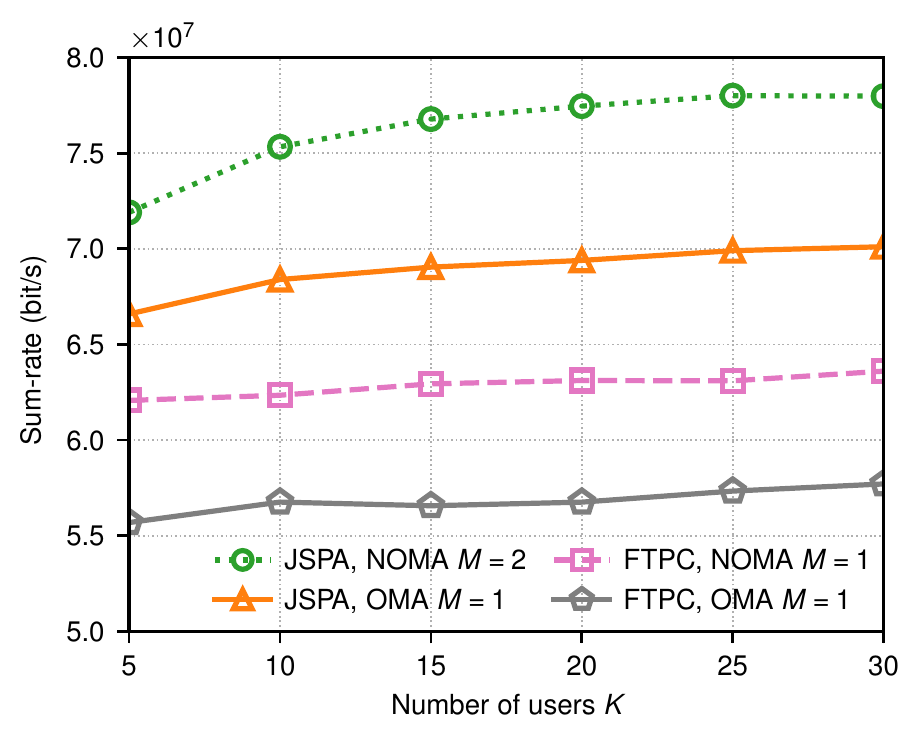}
\vspace{-0.5cm}
\caption{Sum-rate of \textsc{JSPA} and \textsc{FTPC} in a proportional fair scheduler}
\label{fig_PF_WSR}
\vspace{-0cm}
\end{figure}

In Fig.~\ref{fig_PFI} and~\ref{fig_PF_WSR}, we implement a proportional fair scheduler on one frame, which is composed of $T = 20$ time slots. The objective of this setup is to evaluate the fairness performance of our scheme when the weights are chosen according to a proportional fair scheduling. The channel state information is collected at the beginning of the frame. Let $R_k(t)$ be user $k$'s data rate at time slot $t\in\{1,\ldots,T\}$, and $\bar{R}_k(t)$ be user $k$'s average data rate prior to $t$. In the proportional fair scheduling~\cite{girici2010proportional}, $\bar{R}_k(t)$ is updated as follows:
\begin{equation*}
\bar{R}_k(t+1) = \left(1-\frac{1}{T}\right)\bar{R}_k(t) + \frac{1}{T}R_k(t).
\end{equation*}
The weight of user $k$ at time $t$ is then set as $w_k(t) = 1/\bar{R}_k(t)$, in order to achieve a good tradeoff between spectral efficiency and user fairness~\cite{jalali2000data}. At the end of the frame, we average each user $k$'s data rate over the entire frame, i.e., $R_k^{mean} = \sum_{t=1}^T{R_k(t)/T}$. We show respectively in Fig.~\ref{fig_PFI} and~\ref{fig_PF_WSR}, the proportional fairness index defined as $\sum_{k\in\cK}{\log(R_k^{mean})/K}$ and the sum-rate $\sum_{k\in\cK}{R_k^{mean}}$. We compare our scheme \textsc{JSPA} to FTPC with greedy subcarrier allocation. The decay factor of FTPC is set to $0.4$, which is a common value in the literature. Our proposed \textsc{JSPA} outperforms the heuristic FTPC in both user fairness and sum-rate performance, and in both the OMA and NOMA systems. For example, in Fig.~\ref{fig_PF_WSR} for $K=30$, the sum-rate gain of \textsc{JSPA} in NOMA with $M=2$ and in OMA are respectively $23\%$ and $21\%$.

\section{Conclusion}\label{sec:ccl}
We propose a novel approach to solve the WSR maximization problem in MC-NOMA with cellular power constraint. We prove that two sub-problems can be solved optimally with low complexity.
The user selection combinatorial problem is solved by \textsc{SCUS} based on dynamic programming. The multi-carrier power control non-convex problem is solved by the two-stage algorithm \textsc{MCPC}, which uses separability property and gradient descent methods. These algorithms are basic building blocks that can be applied in a wide range of resource allocation problems.
Furthermore, we develop an efficient scheme, \textsc{JSPA}, to tackle the joint subcarrier and power allocation problem. We show through numerical results that it achieves near-optimal WSR and user fairness.

\bibliographystyle{IEEEtran}
\bibliography{IEEEabrv,reference}

\appendix

\subsection{Proof of Lemma~\ref{le:equivalence}}\label{ap:transf}
The objective of~\ref{P} can be written as:
\begin{align*}
&\sum_{k\in\cK}{w_k\sum_{n\in\cN}{R_k^n\left(\bp^n\right)}} = \sum_{n\in\cN}\sum_{k\in\cK}{w_k R_k^n\left(\bp^n\right)}\label{eq:rewrite1}\\
&\stackrel{(b)}{=} \sum_{n\in\cN}W_n\sum_{i=1}^{K}{w_{\pi_n(i)}\log_2\left(\frac{\sum_{j=i}^{K}{p_{\pi_n(j)}^n}+\tilde{\eta}^n_{\pi_n(i)}}{\sum_{j=i+1}^{K}{p_{\pi_n(j)}^n}+\tilde{\eta}^n_{\pi_n(i)}}\right)}\\
&\stackrel{(c)}{=} \sum_{n\in\cN}W_n\sum_{i=1}^{K}{\log_2\left(\frac{\left(\sum_{j=i}^{K}{p_{\pi_n(j)}^n}+\tilde{\eta}^n_{\pi_n(i)}\right)^{w_{\pi_n(i)}}}{\left(\sum_{j=i+1}^{K}{p_{\pi_n(j)}^n}+\tilde{\eta}^n_{\pi_n(i)}\right)^{w_{\pi_n(i)}}}\right)}\\
&\stackrel{(d)}{=} \sum_{n\in\cN}W_n{\left[w_{\pi_n(1)}\log_2\left(\sum_{j=1}^{K}{p_{\pi_n(j)}^n}+\tilde{\eta}^n_{\pi_n(1)}\right)\right.} \nonumber\\
&\phantom{=} + \sum_{i=2}^{K}{\log_2\left(\frac{\left(\sum_{j=i}^{K}{p_{\pi_n(j)}^n}+\tilde{\eta}^n_{\pi_n(i)}\right)^{w_{\pi_n(i)}}}{\left(\sum_{j=i}^{K}{p_{\pi_n(j)}^n}+\tilde{\eta}^n_{\pi_n(i-1)}\right)^{w_{\pi_n(i-1)}}}\right)} \\
&\phantom{=} + \left. w_{\pi_n(K)}\log_2\left(\frac{1}{\tilde{\eta}^n_{\pi_n(K)}}\right) \right].
\end{align*}
Equality~$(b)$ comes from the definition~\eqref{eq:R_2}. At~$(c)$, the weights $w_{\pi_n(i)}$ are put inside the logarithm. Finally, $(d)$ is obtained by combining the numerator of the $i$-th term with the denominator of the $(i-1)$-th term, for $i\in\{2,\ldots,K\}$.

The constant term $w_{\pi_n(K)}\log_2\left(1/\tilde{\eta}^n_{\pi_n(K)}\right)$ can be removed from the objective function, without loss of generality. By applying the change of variables~\eqref{eq:change}, we derive the equivalent problem~\ref{P2}.
Constraints $C1'$ and $C2'$ are respectively equivalent to $C1$ and $C2$ since $x_1^n = \sum_{j=1}^{K}{p_{\pi_n(j)}^n} = \sum_{k\in\cK}{p_k^n}$, for $n\in\cN$. Constraints $C3'$ and $C3''$ come from $C3$ and the fact that $x_i^n - x_{i+1}^n = p_{\pi_n(i)}^n$, for any $i\in\{1,\ldots,K\}$ and $n\in\cN$. In the same way, the active users set in $C4'$ is defined as ${\cU'_n \triangleq \{i\in\{1,\ldots,K\} \colon x_i^n > x_{i+1}^n\}}$. \hfill\ensuremath{\square}


\subsection{Proof of Theorem~\ref{th:algoSCUS}}\label{ap:thalgoSCUS}
\textbf{\textit{Complexity analysis:}} The computational complexity mainly comes from the computation of $V$, $X$ and $U$ in lines~12 to~31. It requires $(M-1)\sum_{j=2}^K{K+1-j} = O(MK^2)$ iterations. Each iteration has a constant number of operations. Thus, the overall worst case computational complexity is $O(MK^2)$.

\textbf{\textit{Optimality:}} We show that, at each iteration of \textsc{SCUS}, the best allocation among $v_0$, $v_1$, $v_2$, is indeed the optimal of~\ref{defV}. As a consequence, $V[M,K,K]$ is by construction the optimal value of~\ref{Psub2}. The corresponding optimal allocation $\bx^n$ can be retrieved from lines~32 to~39.

{\scriptsize$\bullet$} Case $v_0$: Suppose the optimal solution of problem~\ref{defV} is achieved when user $i$ is inactive, we have $x_{i}^n = x_{i+1}^n$ by definition of $\cU'_n$. It follows from $C5'$ and $C6'$ that $x_j^n = \cdots = x_{K+1}^n = 0$. Thus, $V[m,j,i] = V[m,j-1,j-1]$, which is denoted by $v_0$ at line~15.

{\scriptsize$\bullet$} Case $v_1$: Assume that users $i$ and $j-1$ are both active in the optimal solution. Let $x^* = \textsc{MaxF}(j,i, \mathcal{I}^n,\bar{P}^n)$ as in line~14. Suppose for the sake of contradiction that $x^* \geq x_{j-1}^n$. According to Lemma~\ref{le:maxfij}, $f_{j,i}^n$ is increasing in $[0,x^*]$. Hence, the optimal satisfying $C3'$ is achieved for $x_j^n = x_{j-1}^n$, which contradicts with the fact that $j-1$ is active, i.e., $x_{j-1}^n > x_{j}^n$. We derive from $x_{j-1}^n = X[m-1,j-1,j-1]$ that:
\begin{equation}\label{eq:xstar1}
x^* < X[m-1,j-1,j-1].
\end{equation}
Suppose for the sake of contradiction that $x^* = 0$. We know from Lemma~\ref{le:maxfij} that $f_{j,i}^n$ decreases in $\left[x^*,+\infty\right)$. Therefore, the optimal satisfying $C3'$ and $C3''$ is achieved for $x_i^n = 0$, which contradicts with the fact that $i$ is active, i.e., $x_i^n > x_{i+1}^n = 0$. We deduce that: $\hfill x^* > 0. \hfill~\refstepcounter{equation}(\theequation)\label{eq:xstar2}$\\
If~\eqref{eq:xstar1} and~\eqref{eq:xstar2} are satisfied, then we have $V[m,j,i] = v_1$. $v_1$ is the sum of the optimal objective with at most $m-1$ active users from indexes $1$ to $j-1$, i.e., $V[m-1,j-1,j-1]$, and the individual contribution of the active user $i$, i.e., $f_{j,i}^n\left(\textsc{MaxF}(j,i, \mathcal{I}^n,\bar{P}^n)\right) - f_{j,i}^n\left(0\right)$.

{\scriptsize$\bullet$} Case $v_2$: If $i$ is active and $j-1$ is inactive, then $x_{j-1}^n = x_j^n$ by definition of $\cU'_n$. Therefore, $V[m,j,i]$ is given by $v_2 = V[m,j-1,i]$ at line~17.\hfill\ensuremath{\square}

\subsection{Proof of Theorem~\ref{th:algoSCPC}}\label{ap:thalgoSCPC}
\textbf{\textit{Complexity analysis:}} At each for loop iteration $i$, the while loop at line 6 has at most $i$ iterations. Thus, the worst case complexity is proportional to $\sum_{i=1}^{|\cU'_n|}{i} = O(|\cU'_n|^2)$.

\textbf{\textit{Optimality:}} Without loss of generality, we can suppose that the $x_{i_n}^n$'s are initialized to zero. We will prove by induction that at the end of each iteration $i$ at line~10, the following induction hypotheses are true:
\begin{enumerate}[label={$H_\arabic*(i)$:}, ref={$H_\arabic*$}, leftmargin=*, itemindent=3em]
\item ${\sum_{l=1}^{i}{f_{l_n}^n}}$ is maximized by $x_{1_n}^n,\ldots,x_{i_n}^n$,\label{h1}
\item Constraint $C3'$ is satisfied, i.e., $x_{1_n}^n \geq \cdots \geq x_{i_n}^n \geq 0$,\label{h2}
\item For any $l \leq i$, there exists $q \leq l$ and $q' \geq l$ such that variables $x_{q_n}^n,\ldots,x_{q'_n}^n$ are equal and optimal for $f_{q_n,q'_n}^n$. In addition, $f_{q_n,q'_n}^n$ is decreasing for any $y_{q_n}^n,\ldots,y_{q'_n}^n \geq x_{q_n}^n,\ldots,x_{q'_n}^n$ such that $\bar{P}^n \geq x_{q_n}^n \geq \cdots \geq x_{q'_n}^n$.\label{h3}
\end{enumerate}
We set $x_{K+1}=0$, which satisfies $C3''$. $C2'$ is always satisfied since algorithm \textsc{MaxF} gives values in $\left[0,\bar{P}^n\right]$.\\
\textbf{\textit{Basis:}} For $i=1$, $x^*$ computed at line~3 is indeed the optimal of $f_{1_n}^n$. The while loop has no effect since $j = 0 < 1$, therefore $x_{1_n}^n \gets x^*$ and statements~\ref{h1}$(i)$ and~\ref{h2}$(i)$ are true. Since $f_{1_n}^n$ is decreasing for $x_{1_n}^n \in\left[x^*,\bar{P}^n\right]$ according to Lemma~\ref{le:maxfij}, \ref{h3}$(i)$ is also satisfied for $l = 1$ with $q = q' = 1$.\\ 
\textbf{\textit{Inductive step:}} Assume that $x_{1_n}^n(i-1),\ldots,x_{(i-1)_n}^n(i-1)$ are the variables verifying~\ref{h1}$(i-1)$~--~\ref{h3}$(i-1)$ at iteration $i-1<K$. Let the variables at iteration $i$ be $x_{1_n}^n,\ldots,x_{i_n}^n$. 

If by assigning $x^* = \textsc{MaxF}(i_n,i_n, \mathcal{I}^n,\bar{P}^n)$ to $x_{i_n}^n$, it satisfies $C3'$, then ${\sum_{l=1}^{i-1}{f_{l_n}^n(x_{l_n}^n(i-1))}}+f_{i_n}^n(x_{i_n}^n)$ is optimal due to~\ref{h1}$(i-1)$ and the fact that $x_{i_n}^n$ is maximal for $f_{i_n}^n$. The iteration terminates with $x_{i_n}^n \gets x^*$ and $x_{l_n}^n = x_{l_n}^n(i-1)$ for all $l < i$. Thus, \ref{h1}$(i)$ and~\ref{h2}$(i)$ are satisfied and~\ref{h3}$(i)$ is true for index $l = i$ with $q= q' = i$.

Otherwise if $x_{i_n}^n = x^*$ violates $C3'$, we have $x_{j_n}^n(i-1) < x^*$ at the while loop's first iteration at line~6 with $j \gets i-1$. We deduce from the monotonicity of $f_{l_n}^n$, $l\leq i$, that:
\begin{equation}\label{proof:violates}
x_{i_n} \geq x_{j_n}^n(i-1).
\end{equation}
We know from~\ref{h3}$(i-1)$ there exists $q \leq j$ such that $x_{q_n}^n(i-1),\ldots,x_{j_n}^n(i-1)$ are equal and optimal for $f_{q_n,j_n}^n$ and that $f_{q_n,j_n}^n$ is decreasing for any increase in these variables. It follows from Eqn.~\eqref{proof:violates} that the optimal is reached at $x_{q_n}^n = \cdots = x_{j_n}^n = x_{i_n}$.

As a consequence, the maximum of ${\sum_{l=q}^{i}{f_{l_n}^n}}=f_{q_n,i_n}^n$ is given by $x^* \gets \textsc{MaxF}(q_n,i_n, \mathcal{I}^n,\bar{P}^n)$.
Then, the algorithm sets $x_{q_n}^n,\ldots,x_{i_n}^n \gets x^*$ and $x_{l_n}^n = x_{l_n}^n(i-1)$ for all $l < q$. ${\sum_{l=q}^{i}{f_{l_n}^n(x_{l_n}^n)}}$ is optimal by construction. ${\sum_{l=1}^{q-1}{f_{l_n}^n(x_{l_n}^n)}}$ remains unchanged and optimal by induction hypothesis at iteration $q-1$. Therefore,~\ref{h1}$(i)$ is true. Property~\ref{h3}$(i)$ remains unchanged on $l <q$, and is also true for any $l \in\{q,\ldots,i\}$ with indexes $q$ and $q' = i$. The only property that may not be satisfied at this point is~\ref{h2}$(i)$, since $x_{(q-1)_n}^n < x^*$ is possible.

We continue this reasoning until reaching the highest index $j \in\{2,\ldots,i-1\}$ such that $x_{(j-1)_n}^n \geq x^* = \textsc{MaxF}(j_n,i_n, \mathcal{I}^n,\bar{P}^n)$. We know that this index exists since all variables are upper bounded by $\bar{P}^n$ and $x_{1_n}^n = \bar{P}^n$ due to Lemma~\ref{le:maxfij}. Properties~\ref{h1}$(i)$ and~\ref{h3}$(i)$ remain true by the above construction. Finally,~\ref{h2}$(i)$ is satisfied at this point. \null \hfill\ensuremath{\square}

\subsection{Proof of Lemma~\ref{lemma:Fconcave}}\label{ap:Fconcave}
The convexity of $\mathcal{F}$ is direct from its definition in Eqn.~\eqref{eq:feasibleSet}. We now prove that each $F^n$ is concave with respect to $\bar{P}^n\in\left[0,P_{max}^n\right]$, $n\in\cN$. Let $x_{1_n}^n,\ldots,x_{|\cU'_n|_n}^n$ be the allocation returned by \textsc{SCPC}($\mathcal{I}^n, \cU'_n, P_{max}^n$). We can see in Algorithm~\ref{algoSCPC}, that the only impact of $\bar{P}^n$ on the allocation is through \textsc{MaxF} thresholding (see lines~4 and~6 in Algorithm~\ref{algo:MaxF}). Hence, the allocation returned by \textsc{SCPC}($\mathcal{I}^n, \cU'_n, \bar{P}^n$) is equal to $\min\{x_{1_n}^n,\bar{P}^n\},\ldots,\min\{x_{|\cU'_n|_n}^n,\bar{P}^n\}$.
Besides, the variables can be divided in consecutive groups of the form $x_{q_n}^n,\ldots,x_{q'_n}^n$, such that $x_{q_n}^n = \cdots = x_{q'_n}^n$ is optimal for $f_{q_n,q'_n}^n$, according to~\ref{h3} of Theorem~\ref{th:algoSCPC}'s proof. Lemma~\ref{le:maxfij} implies that $f_{q_n,q'_n}^n$ is concave on $\left[0,x_{q_n}^n\right]$, therefore $f_{q_n,q'_n}^n\left(\min\{x_{q_n}^n,\bar{P}^n\}\right)$ is concave in $\bar{P}^n\in\left[0,P_{max}^n\right]$. It follows by summation that $F^n$ and $\sum_{n\in\cN}{F^n}$ are concave, which completes the proof.
\null \hfill\ensuremath{\square}

\end{document}